# A New Index for Quantifying Stability Adaptation Effort in Power Electronics-Dominated Power Systems Based on Frequency-Domain Impedance Identification

**Luis A. Garcia-Reyes[1,2]*, Macarena Martín-Almenta[1], Edgar Nuño-Martínez[1], Javier Renedo[1], Antonio Cordón-Rodríguez[1], Oriol Gomis-Bellmunt[2], Eduardo Prieto-Araujo[2]**
**[1]Red Eléctrica – Redeia, Spain**
**[2]CITCEA – UPC, Spain**
**{Luis.GarciaR*, Macarena.Martin, Edgar.Nuno, Javier.Renedo, acordon}@ree.es**
**{oriol.gomis, eduardo.prieto-araujo}@upc.edu**

**SUMMARY**

The increasing use of converter-based resources has made stability assessment in modern power systems more challenging, particularly because many devices are available only as black-box models. Conventional eigenvalue-based methods require full dynamic representations, while frequency-domain impedance techniques based on the Generalized Nyquist Criterion (GNC) provide stability insight but do not quantify how much a converter must be retuned to restore stability. This paper introduces the Adaptation Effort Index (AEI), a metric that measures the sensitivity of dominant modes to parameter variations and quantifies the effort required for a converter to reconfigure itself under unstable operating conditions. The AEI combines the Nyquist Stability Margin obtained from frequency-domain screening with a modal-sensitivity index derived from fitted state-space models synthesized from black-box impedance data. The method is applied to a system comprising a grid-forming and a grid-following converter. Results show that, although both units can be tuned to stabilize the system, the grid-following converter requires significantly less adaptation effort. The AEI therefore provides a practical and model-agnostic tool for identifying which converter should be reconfigured in converter-dominated grids, offering valuable guidance for system operators and manufacturers.

# 1 Introduction

The increasing penetration of power electronics in modern power systems, driven by the massive integration of renewable energy sources, has significantly altered system dynamics. This shift has introduced new stability and interaction challenges arising from the lack of inertia, the complexity of converter control schemes, and the high switching frequencies characteristic of power electronic converters [1, 2]. These issues must be thoroughly assessed, particularly by transmission system operators (TSOs), who are responsible for ensuring the stable and continuous operation of the grid, especially when new converter-based resources (CBRs) are integrated. To guarantee such stability, newly connected CBRs are required to adapt and reconfigure their control systems in accordance with local grid codes to avoid adverse interactions and instability with the network and with neighboring CBRs [3, 4].

However, as illustrated in cases such as the Balearic Islands, the stability and interaction challenges associated with increasing power transfer through a high-voltage direct current (HVDC) link, implemented using voltage-source converter (VSC) technology in addition to an existing line-commutated converter (LCC), have been investigated in [5]. The study identified a risk of commutation failure in the existing HVDC link but did not provide detailed mitigation strategies. This raises a critical question for both system operators and manufacturers: which unit is more capable of adapting and reconfiguring under a specific scenario where stability cannot be guaranteed—the newly installed CBR or the existing neighboring CBR. Furthermore, from an industrial perspective, CBR control design and tuning are subject to operational requirements and technical limits that manufacturers must respect to ensure proper device operation in accordance with grid codes [6].

Conventional small-signal tools and methodologies can be used to analyze the emerging stability issues introduced by CBRs [7]. However, their applicability is limited because CBR projects are typically developed by multiple manufacturers who impose intellectual property (IP) restrictions on their models, resulting in power systems dominated by "black-box" representations [2, 8]. This lack of access to internal CBR models poses significant challenges for stability assessment, particularly for TSOs. As demonstrated in [5], the absence of detailed models makes it difficult to propose effective mitigation measures.

Small-signal methods based on eigenvalue analysis (modal analysis) are widely used to study stability problems associated with CBR integration [2]. Nevertheless, they require detailed dynamic models of the system—information that is often unavailable in modern black-box-dominated grids. In contrast, small-signal impedance-based methods can assess CBR-driven stability issues in black-box environments because they rely on open-loop frequency-domain identification and apply closed-loop stability criteria to determine system stability [9, 10]. Although robust and increasingly accepted, impedance-based methods are limited to analyzing a single point in the system and do not reveal which specific components contribute to instability, as they lack explicit state variables that can be associated with system dynamics—unlike modal analysis [8, 11]. Consequently, both methodologies face limitations when performing a comprehensive stability assessment of black-box-dominated systems and, therefore, when extracting meaningful metrics to determine which CBR should more readily adapt and reconfigure under a given instability scenario.

To address this gap, this paper introduces a metric that quantifies the adaptability of a new CBR when connected to a grid that already hosts an existing CBR, with a focus on modern power systems characterized by black-box models. The proposed metric combines the frequency-domain stability margin, computed using the Generalized Nyquist Criterion (GNC), with a new index that estimates the maximum dynamic effort a CBR can exert. This index is derived by identifying dominant modes through a black-box state-space synthesis methodology previously presented in [8]. The proposed metrics are applied to a system composed of two VSCs connected to a grid equivalent: one operating in grid-forming (GFM) mode and the other in grid-following (GFL) mode.

## 2 Proposed Approach for the Adaptation Effort Index (AEI)

The Adaptation Effort Index (AEI) quantifies the dynamic effort experienced by an electrical system when an operational parameter is modified, directly evaluating the sensitivity of system modes [12] in both the frequency-domain and the modal (state-space) domain. The metric is grounded in the observation that dominant modes, those exerting the greatest influence on stability and transient behavior, can shift significantly under parametric variations. Such shifts provide early indications of risk and allow the effort required to maintain stability to be quantified even before a formal loss of stability occurs. Therefore, the proposed index is intended for pre-connection impact studies, as shown in Figure 1, where, after performing a screening impact analysis, the AEI can provide insight into the effort required to maintain the stability of the new CBR to be connected.

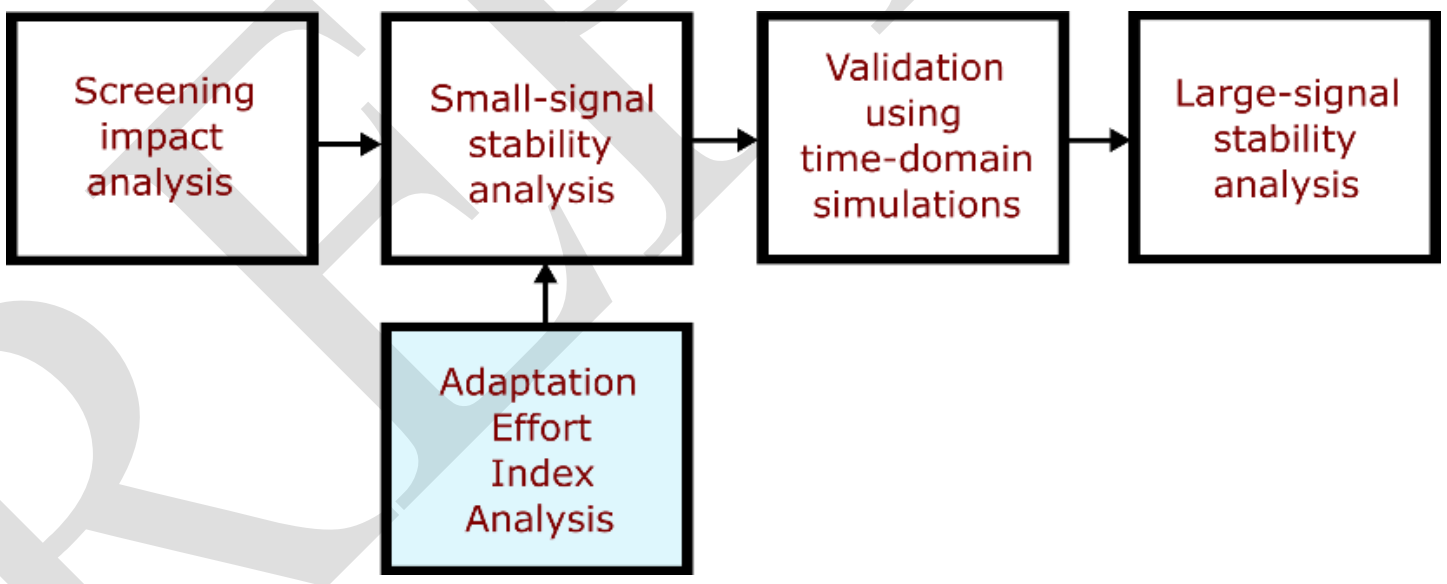

***Figure 1 – Stability analysis workflow for new CBRs, based on the scheme proposed in [2].***

The AEI is constructed using two complementary methodologies. The first computes the minimum distance between the critical point (-1,*j*0) and the trajectory of the closed-loop eigenvalues obtained from the Nyquist plot. The second relies on identifying the dominant mode using the function `fitss` for black-box state-space synthesis presented in [8], from which the maximum change in the dominant mode required to reconfigure the system toward stability is extracted. Both methodologies are briefly described below.

### 2.1 Nyquist Stability Margin (NSM)

Impedance-based methods rely on frequency scanning to characterize impedance or admittance matrices in the frequency-domain (FD). This is achieved by injecting small sinusoidal voltage or current perturbations at specific frequencies using series voltage sources or parallel current sources. These perturbations are typically kept within 1–3% of the nominal voltage or current to preserve steady-state operation and maintain the system linear behavior [9]. In this work, the SIaD-Tool [9], an open-access tool for FD stability and interaction analysis, is employed to accurately identify the matrices $Y_{\text{sys1}}$ and $Z_{\text{sys2}}$ required for impedance-based analysis of two interconnected systems, as illustrated in Figure 2.

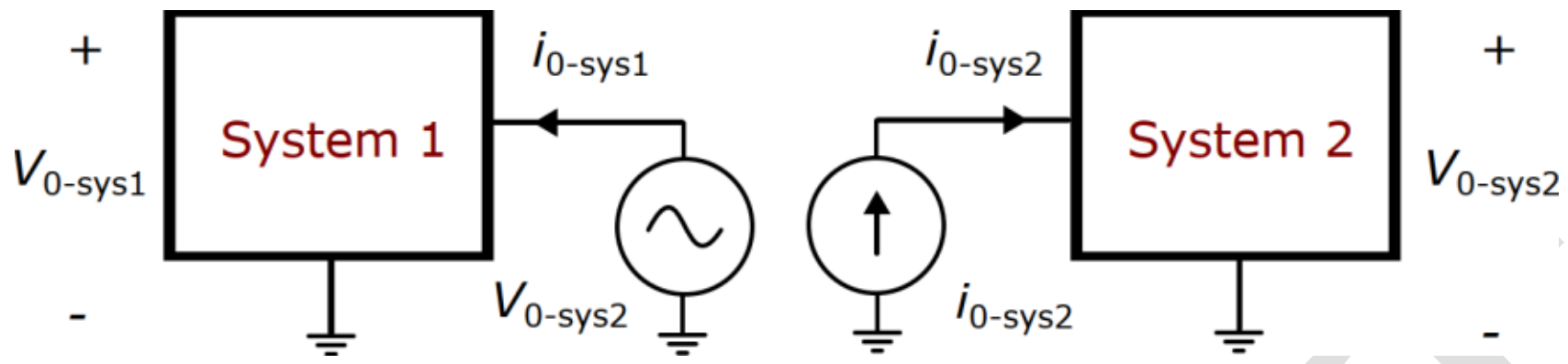

***Figure 2 – Equivalent decoupled system for impedance-based analysis.***

Here, $Y_{\text{sys1}}$ corresponds to the new CBR being connected, while $Z_{\text{sys2}}$ represents the existing system. Once both matrices have been identified, the closed-loop transfer matrix is given by:

$$T(j\omega) = \left[I + Y_{sys1}(j\omega)Z_{sys2}(j\omega)\right]^{-1} Y_{sys1}(j\omega), \tag{1}$$

where the minor loop is defined as:

$$L(j\omega) = I + Y_{sys1}(j\omega)Z_{sys2}(j\omega). \tag{2}$$

The Generalized Nyquist Criterion (GNC) evaluates the trajectory of $L(j\omega)$ and states that stability is ensured if this trajectory does not encircle the critical point (-1,j0) in the Nyquist plane [13]. The Nyquist Stability Margin (NSM) quantifies the minimum distance between the closest point of the trajectory and the critical point (-1,j0) for each operating point $k$, providing a numerical measure of proximity to instability. The NSM is defined as:

$$NSM(k) = \min_{\omega} \left|\det(L(j\omega, k))\right|. \tag{3}$$

According to the results presented in [9], a small NSM indicates that the system operates close to the critical point, whereas a large NSM reflects a robust system that requires a greater effort to be driven out of its stable operating region.

### 2.2 AEI Based on Fitted State-Space (fitss)

The `fitss` algorithm, introduced in [8], enables the synthesis of a state-space model for a black-box system using the impedance and admittance matrices identified in the FD. This is achieved through a vector-fitting-based procedure [14] combined with model-order reduction techniques [15] and the pole-expansion strategy introduced in [8]. To obtain the Multiple Input Multiple Output (MIMO) state-space representation of the system it is necessary to fit independently each Single Input Single Output (SISO) sub-system and then assemble the reduced models accordingly to the input-output relationships (voltage and current), depending on whether admittance or impedance data are used. The algorithm view is given in Algorithm 1 (Appendix) and is described in detail in [8], including full benchmark validation of the reconstructed model. For the initial order is proposed a low order between 2 and 4, since the pole-expansion strategy tends to adapt the order according to the root-mean-square (RMS) error. The `fitss` approach reconstructs the system dynamic behavior with an average modal error of approximately 1.5% [8]. The resulting black-box state-space model takes the form:

$$\begin{aligned} \Delta\dot{x} &= A\Delta\mathrm{x} + \mathrm{B}\Delta\mathrm{u}, \\ \Delta\mathrm{y} &= \mathrm{C}\Delta\mathrm{x} + \mathrm{D}\Delta\mathrm{u}, \end{aligned} \tag{4}$$

where x denotes the states and u and y denote the inputs and outputs, respectively. This representation lacks explicit meaning for the states but provides direct access to the system modes through the characteristic equation:

$$det(\lambda_i I - A) = 0, \tag{5}$$

where $\lambda_i = \sigma_i \pm j\omega_i$ denotes the $i$-th mode of the system, with $\sigma_i$ representing the damping and $\omega_i$ the angular frequency in rad/s. For each operating point, all system modes are identified, and the dominant mode $\lambda_{dom}$, is selected as the one with the largest real part:

$$\lambda_{dom} = \underset{i}{\mathrm{argmax}}\big(\Re(\lambda_i)\big). \tag{6}$$

Because the dominant mode may change identity across operating points, a modal-tracking procedure based on minimum distance in the complex plane is implemented. This ensures that each modal branch is followed consistently throughout the parameter sweep, preventing spurious jumps or incorrect mode assignments.

Let $x$ be the operational parameter being varied (e.g., a control-loop gain). For each operating point $k$, the derivative of the dominant mode with respect to this parameter is computed using centered finite differences [16]. This derivative captures the rate at which the dominant mode moves in the complex plane as the parameter changes. The maximum rate of change yields the largest displacement of an eigenvalue per unit variation of the parameter, defining the AEI as:

$$AEI = \max_k \left|\frac{\partial\lambda_{dom}(k)}{\partial x}\right|, \tag{7a}$$

$$AEI_{pu} = \max_k \left|\frac{k_{base}}{\lambda_{base}} \cdot \frac{\partial\lambda_{dom}(k)}{\partial x}\right|, \tag{7b}$$

where $k_{base}$ and $\lambda_{base}$ are the base values for the parameter change ratio and the dominant mode, respectively. This index quantifies the dynamic effort experienced by the system: large values indicate that small parameter variations produce significant modal shifts, revealing high sensitivity. The AEI is evaluated for both the real and imaginary components of the dominant mode to determine whether the effort is primarily associated with damping or with oscillatory frequency. The AEI procedure can be applied independently to each dominant mode present in the system, producing mode-wise effort measures that can be compared to prioritise corrective actions.

The AEI can be used as a sufficient (conservative) stability condition to guarantee that the dominant mode does not cross the imaginary axis for a parameter variation $\Delta x$. Let $\lambda_0 = \lambda_{dom}(k_0)$ denote the dominant eigenvalue at operating point $k_0$, then the condition is:

$$\Re(\lambda_0) + AEI \cdot |\Delta x| < 0, \tag{8}$$

where the maximum permissible parameter displacement is given by:

$$\Delta x_{max} = -\Re(\lambda_0) \, / \, AEI. \tag{9}$$

Here a safety factor $h > 1$ may be introduced to account for identification error and nonlinearities, requiring $|\Delta x| < \Delta x_{max}/h$.

Finally, using (8)-(9) together with (3) provides a practical acceptance criterion: a new CBR is accepted at a given operating point only if (8) holds for the expected $\Delta x$ with the chosen safety margin and the corresponding NSM exceeds the minimum threshold.

# 3 Study Case: Grid-Connected Grid-Forming and Grid-Following

To validate the proposed effort metric and to illustrate its applicability and interpretability, a system composed of two grid-connected converters is considered: one operating in GFM mode and the other in GFL mode, as shown in Figure 3. Both converters are black-box models, meaning that their detailed internal representations are not accessible. However, a state-space representation of the Thevenin equivalent of the grid is available. The system parameters used in this study are summarized in Table 1, where the Thevenin equivalent exhibits an X/R ratio of 10 and a short-circuit ratio (SCR) of 10, and the connected load is 66.6 MW. Note that the methodology is aplicable to linear models and that, for each operating point (SCR or X/R variation), a different fitss model is derived in accordance with modal-analysis theory [17].

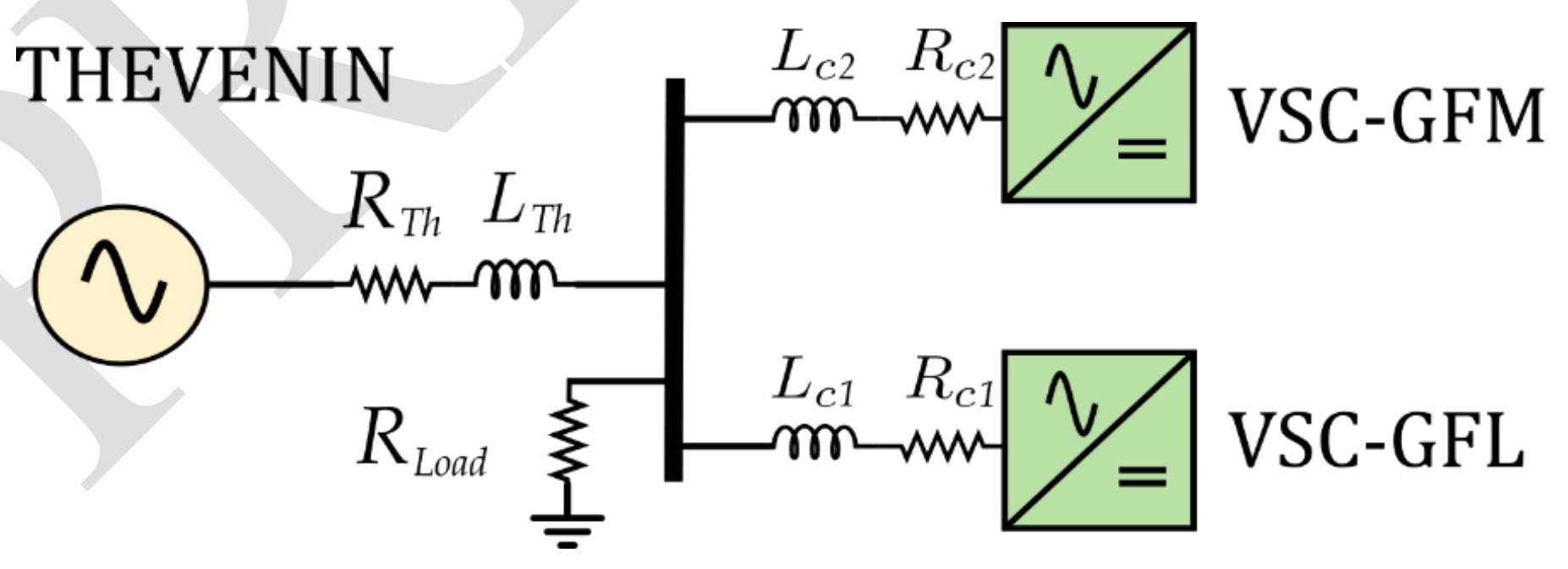

*Figure 3 – One-line diagram of the test network.*

*Table 1 – Parameters for the GFL and GFM converters.*

| Parameter | GFL | GFM | Unit |
|---|---|---|---|
| $S_{base}$ | 100.0 | 100.0 | MVA |
| $V_{base}$ | 230.0 | 230.0 | kV |
| $V_{DC}$ | 400.0 | 400.0 | kV |
| $R_C$ | 1.058 | 1.058 | Ω |
| $L_C$ | 0.118 | 0.118 | H |
| $K_{pPLL}$ | 10E-5 | - | 1/s |
| $K_{iCC}$ | 1058 | 1058 | V/A·s |
| $K_{pCC}$ | 117.9 | 117.9 | V/A |
| $K_{droop\text{-}U}$ | 6.2E-4 | 10E3 | pu/pu |
| H | - | 1 | s |

In this study case, the integration of the GFM converter into the system is analyzed. After tuning the parameters accessible within the black-box model (conventional tuning), system-wide stability cannot be achieved. Figure 4 shows the time-domain instantaneous current response through the GFM link for the initial tuning setup. Instability is confirmed when, at 5 s, the gain $k_{\text{droop-U}}$ in the reactive-power and voltage (Q-U) droop controller of the GFM is activated. This gain is the parameter selected for the analysis and application of the proposed methodology in the test system; however, the procedure is applicable to any accessible parameter in black-boxed systems.

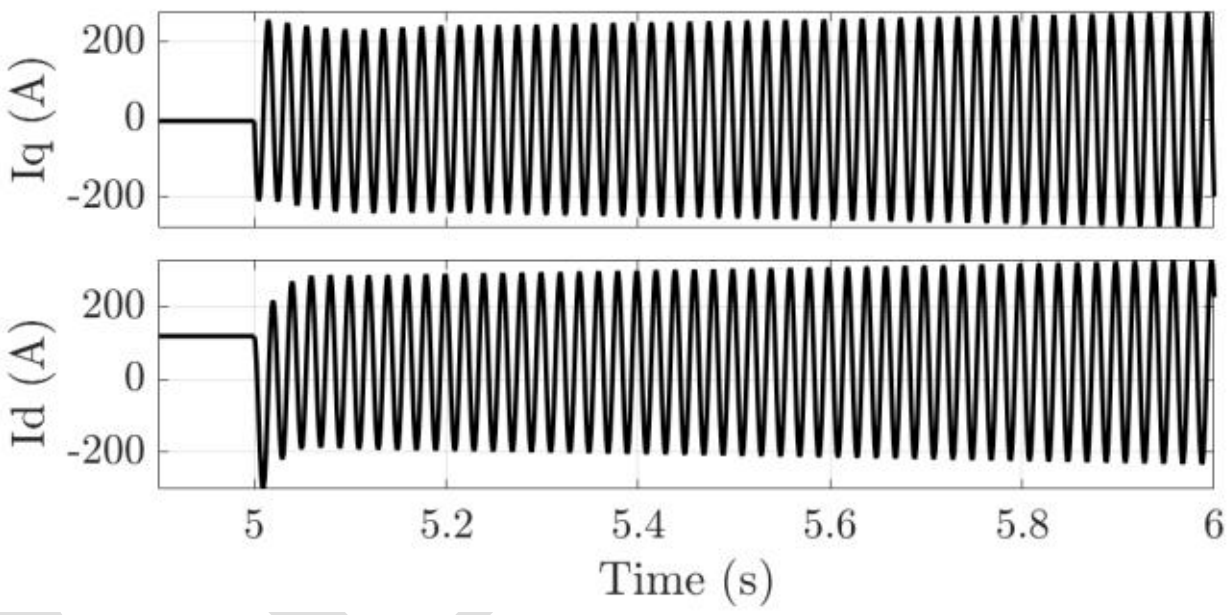

***Figure 4 – Unstable instantaneous current in the GFM link after activation of the original*** $k_{droop\text{-}U}$ .

Consequently, a screening study is performed using the SIaD-Tool to obtain the detailed GNC of the closed-loop system (i.e., the eigenvalues of $L(j\omega)$) using 200 linearly spaced points over a frequency range of 1–2000 Hz. The time window used for the scanner was 1 s that means a resolution of 1 Hz in the frequency-domain and a time step of 50 µs. This small-signal procedure is standard practice when a new device is connected to the grid [2].

The results of the screening study are shown in Figure 5, where, for simplicity, only eigenvalue 1, the one involved in the instability, is displayed. The plot shows that this eigenvalue performs an encirclement when the original value of $k_{\text{droop-U}}$ is used. However, when this gain is reduced by approximately 80%, the trajectory of eigenvalue 1 increases its NSM and crosses below the critical point without producing any encirclements (see the detailed view in Figure 5b).

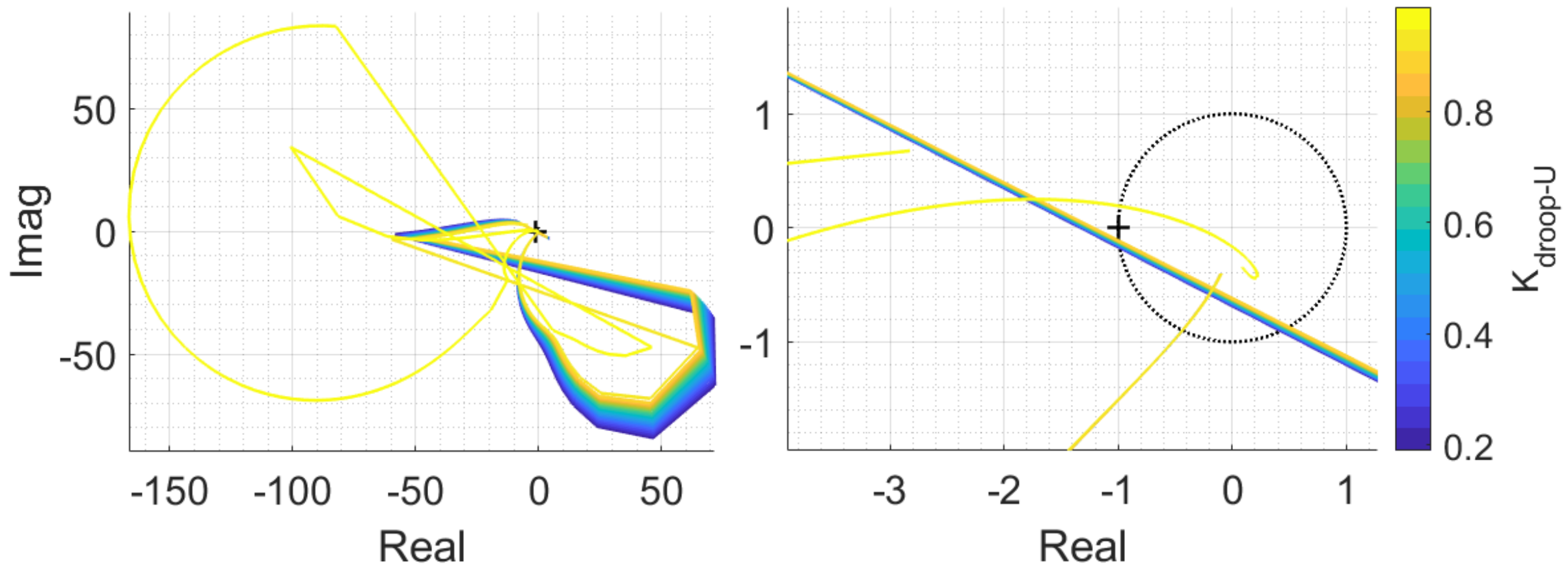

*a)* *b)*

***Figure 5 – GNC results for the GFM grid-connected: a) Nyquist plot and b) detail in the critical point.***

Adjusting the GFM $k_{\text{droop-U}}$ control-loop gain is sufficient to restore system-wide stability, as shown in Figure 6 where the instantaneous current becomes stable after 5 s when the $k_{\text{droop-U}}$ is activated with a reduction of 80%. However, to determine whether it requires less effort for the GFM converter to reconfigure itself compared to its GFL counterpart, an AEI-based analysis is performed to identify which unit has a greater ability to adapt in order to stabilize the overall system. This analysis is divided into two parts. First, the AEI of the GFM is evaluated when its $k_{\text{droop-U}}$ gain is reduced by 70%, while keeping the GFL gain unchanged. Second, the AEI of the GFL is assessed when its corresponding gain is reduced by the same proportion, with the GFM gain left unmodified. The resulting AEI values for both cases are presented below.

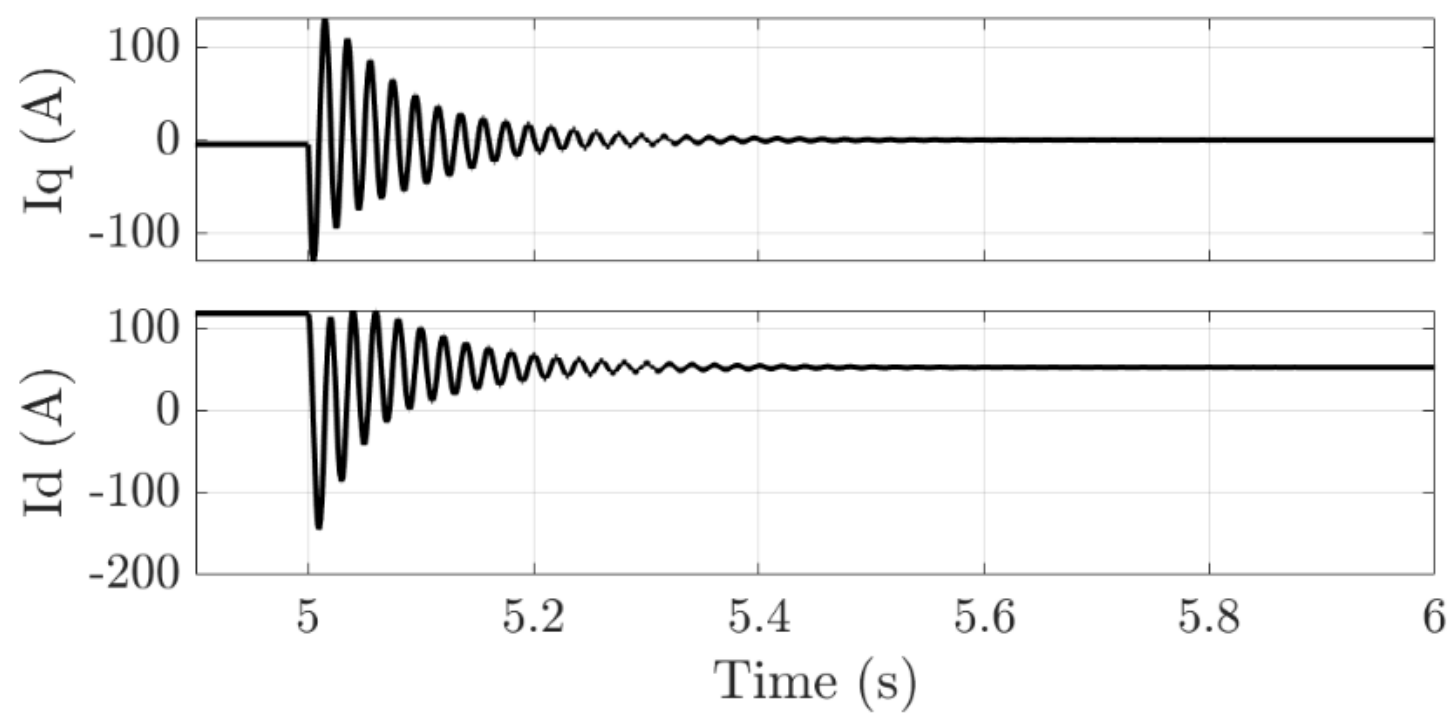

***Figure 6 – Stable instantaneous current in the GFM link after activation of the 80% of $k_{droop\text{-}U}$ .***

### 3.1 AEI for the Grid-Forming Converter

The SIaD-Tool is used to identify the black-box GFM and GFL models at each operating point resulting from reducing the GFM $k_{\text{droop-U}}$ gain by up to 70%. The `fitss` algorithm is then applied to obtain a state-space model for each operating point of both converters, using an initial order of 4 with complex poles uniformly distributed across the same frequency band and using the same sampling resolution in both frequency and time. The inputs and outputs are selected to be consistent with the connection scheme, and a connect function is used to combine the fitted state-space representations of the converters with the Thevenin equivalent. The function connects the dynamic state-space systems based on the input and output channels with matching the names, delivering a complete state-space representation. Here, the full small-signal model is obtained using the open-source tool STAMP [18]. Figure 7 presents the time-domain validation of the fitss model for the stable and unstable cases after applying a 5% voltage step at the grid side at 7 s, assuming the activation of the $k_{\text{droop-U}}$ at 5 s. As shown, for the unstable case fitss cannot capture neither the initial steady state nor the transient response of the time-domain simulation since the FD identification is not stable leading to a state-space model automatically no stable.

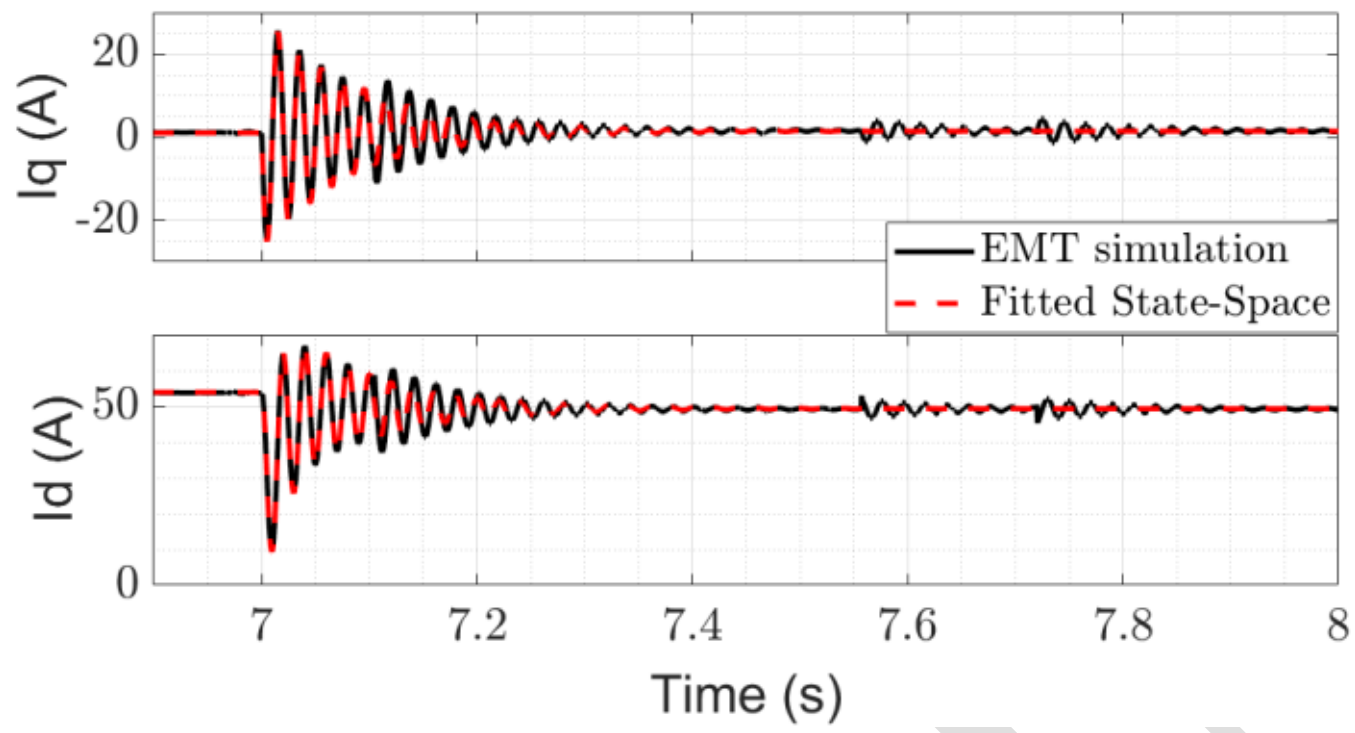

***a)***

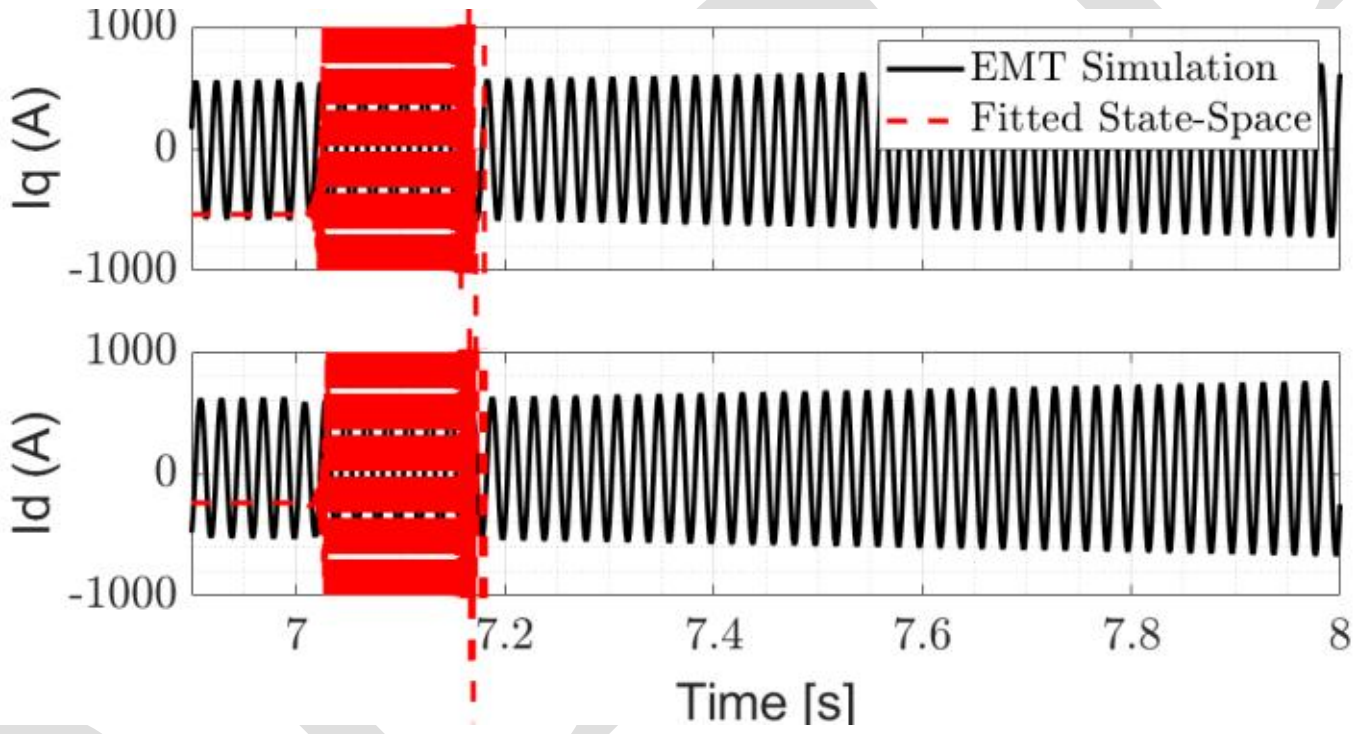

***b)***

***Figure 7 – Fitted state-space validation for the GFM VSC: a) stable and b) unstable case.***

Figure 8 shows the modal results from the sensitivity study performed by varying the GFM $k_{\text{droop-U}}$ gain. Only a single mode is involved in the instability in this test case, located at $277 \pm j6165$, corresponding to the operating point with 100% of the original GFM gain. As this gain is reduced, the unstable mode shifts significantly to the left in the complex plane, and no other mode crosses into the right-half plane, thereby restoring stability to the overall system.

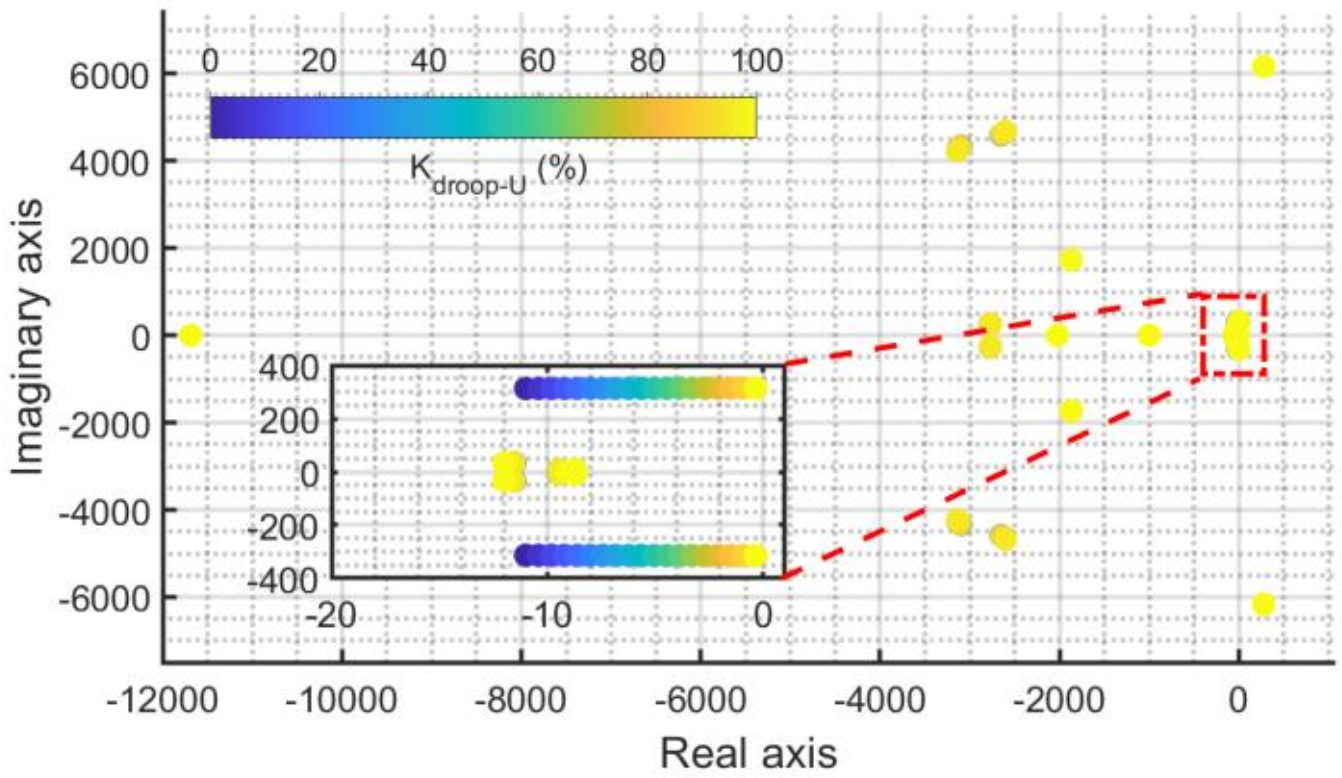

***Figure 8 – Eigenvalue sensitivity in the test system when varying the GFM $k_{droop\text{-}U}$ gain.***

Figure 9 presents the NSM evaluated at each operating point. The NSM remains approximately constant at around 0.75 as the gain is reduced. However, at the nominal 100% gain, the NSM drops to nearly 0.2, indicating that the determinant of the GNC approaches the critical point, making the system extremely susceptible to instability. Given the magnitude of this reduction, the distance suggests practical instability, consistent with the behavior observed in Figure 5 from the detailed GNC screening.

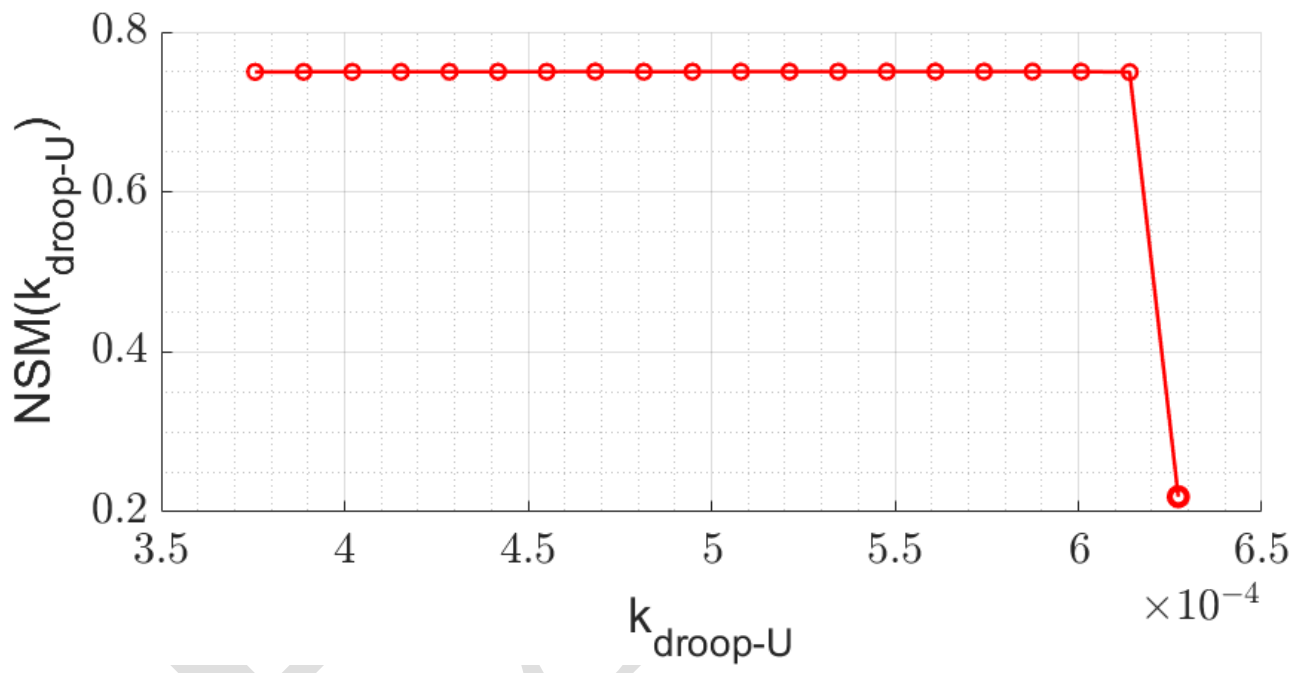

***Figure 9 – NSM index when varying the GFM $k_{droop\text{-}U}$ gain.***

Figure 10 shows the quantitative results for the dominant mode and the AEI across operating points. The computed and normalized AEI is 18.03, a value that reflecs the abrupt change in the system stability conditions. The dominant mode is primarily influenced by its oscillatory component as shown in Figure 10a, although it also exhibits a significant damping contribution. As shown in Figure 10a, the damping remains relatively constant across parameter variations. At the beginning of the sweep and near the stability boundary, the AEI experiences a sharp increase as shown in Figure 10b. This highlights the "stress" the system undergoes for parameter variations between $3.5 \times 10^{-4}$ and $4.5 \times 10^{-4}$ and again beyond $6 \times 10^{-4}$. Outside those ranges the system experiences less effort, so stability is reached more easily and with lower "stress." This is important because the method applies to any accessible controller parameter; the same analysis can identify parameter ranges where the system adapts more readily and with lower stress.

Finally, checking the local stability condition (8) with a safety factor of 2 across all operating points shows that only the operating point corresponding to 100% of the original gain fails the condition. This indicates that, at that operating point, the GFM cannot be connected without risking instability. This conclusion complements the results in Figure 9, where the NSM drops abruptly at that operating point.

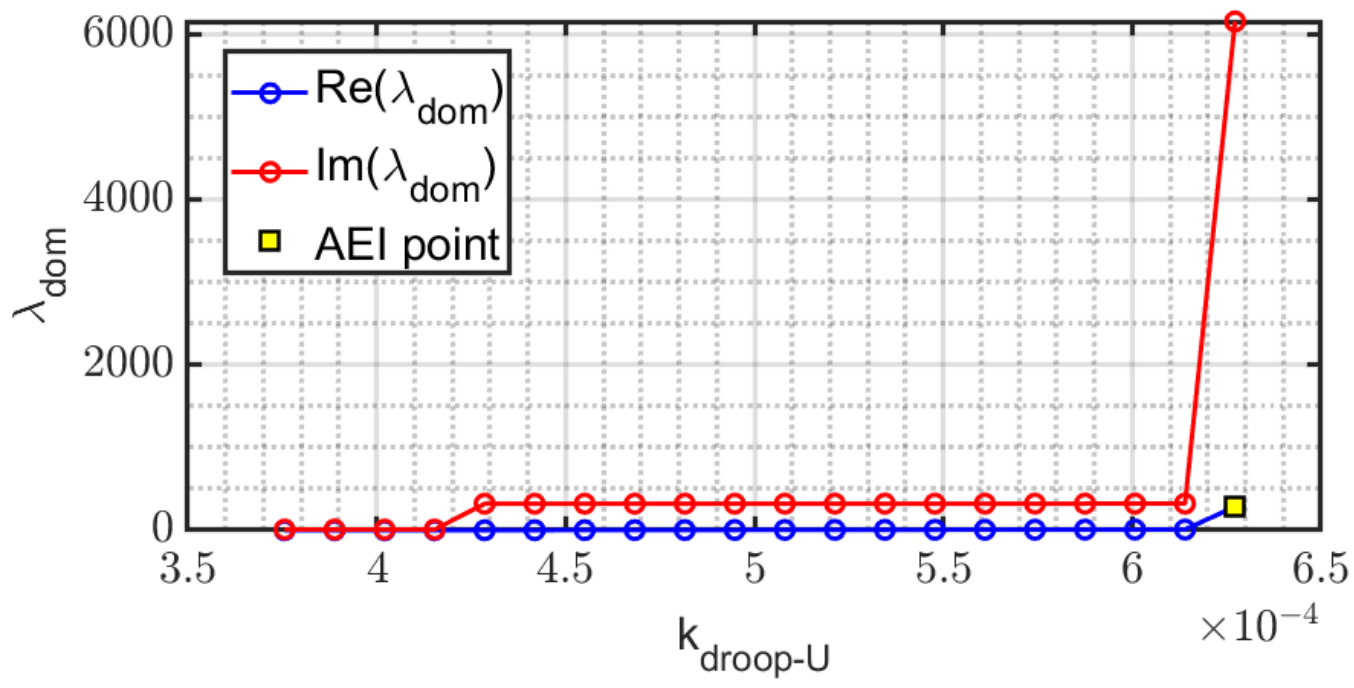

*a)*

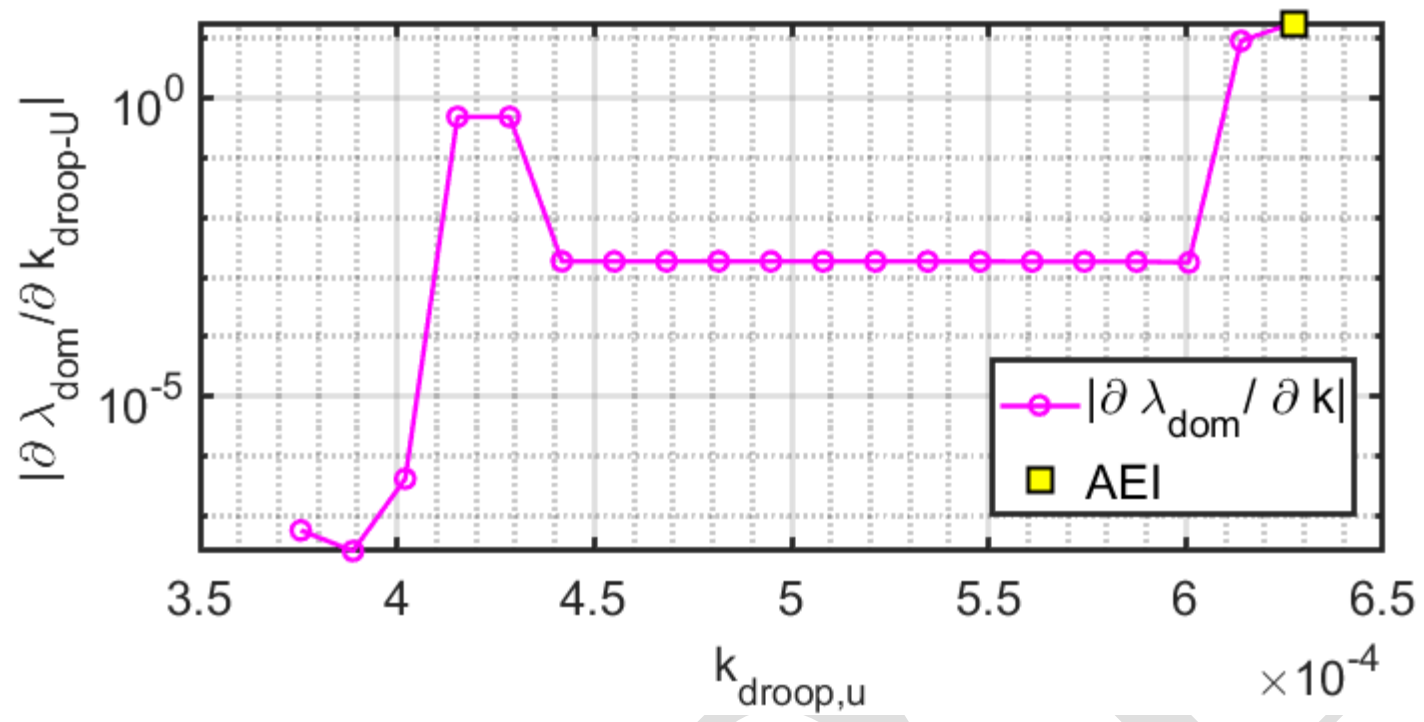

*b)*

***Figure 10 – Adaptation index results: a) dominant mode behaviour and b) AEI.***

### 3.2 AEI for the Grid-Following Converter

Using the same identification and state-space synthesis framework applied to the GFM case, the procedure is now repeated to identify the operating points obtained by varying the reactive-power and voltage droop gain of the GFL, while keeping the GFM gain at its original value. To ensure a fair comparison and avoid misleading conclusions regarding the relative adaptation effort of the GFL versus the GFM, the GFL $k_{\text{droop-U}}$ gain is reduced by the same proportion as in the previous case, namely, 70%. This is precisely where the AEI becomes valuable, as it quantifies how much the dominant mode shifts per unit change of the parameter.

Figure 11 shows the sensitivity analysis results when the GFL is reconfigured to restore full system stability, with its gain reduced from 100% to 30%. The unstable mode is nearly identical to the one identified in the previous GFM case, with a variation of approximately 1.2%. Notably, when the droop gain is reduced, the next dominant mode lies much closer to the original one, resulting in a significantly smaller rate of change, and therefore a lower adaptation effort, to achieve stability.

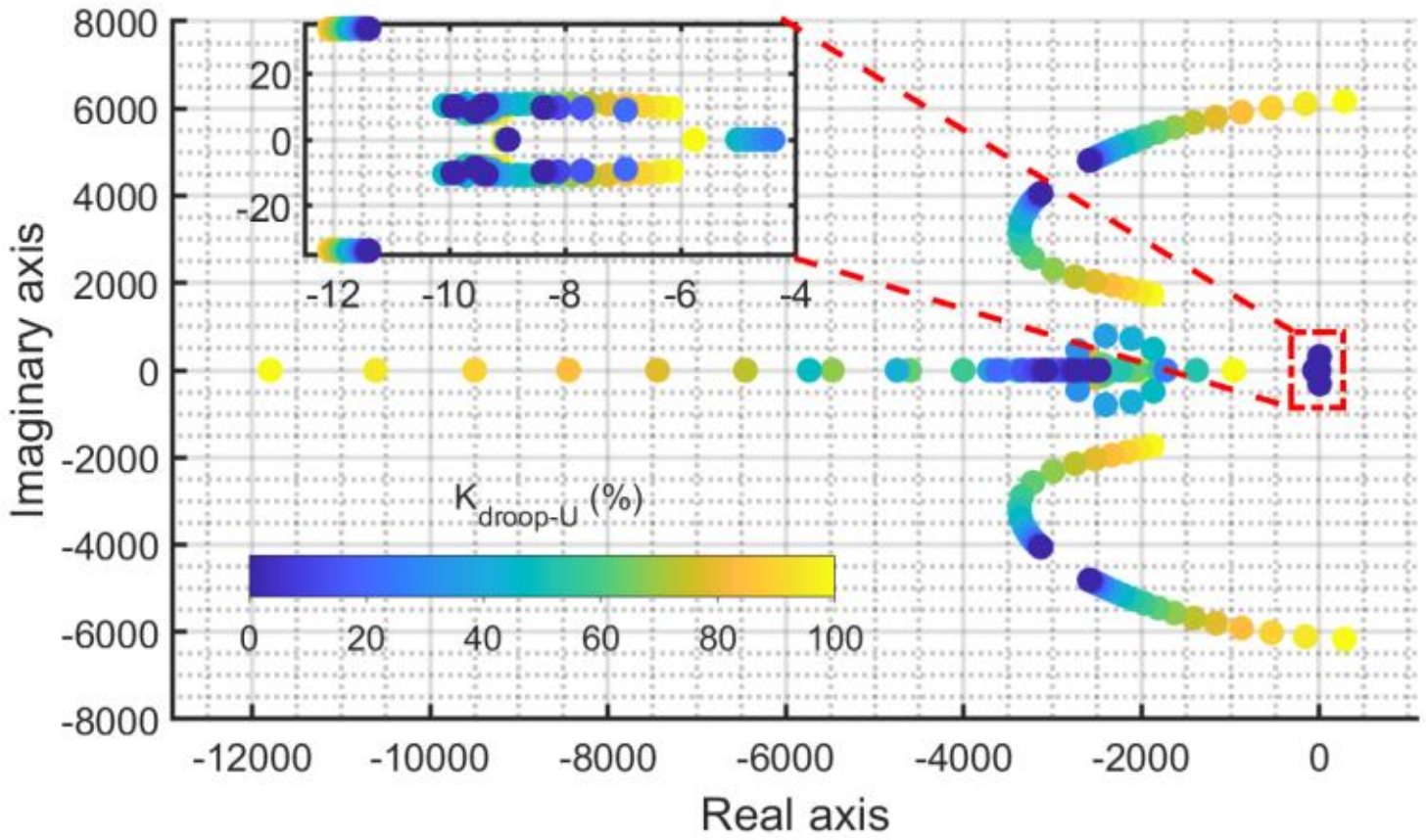

***Figure 11 – Eigenvalue sensitivity in the test system when varying the GFL $k_{droop\text{-}U}$ gain.***

Figure 12 presents the NSM results for each operating point. When the gain is reduced by 70%, the NSM becomes very large, indicating strong robustness and stability. However, as the gain increases, the NSM drops below 0.3. This behavior is consistent with the dominant-mode evolution shown in Figure 13a, which exhibits a gentle slope indicating a constant rate of change compared to the GFM case shown in Figure 10a.

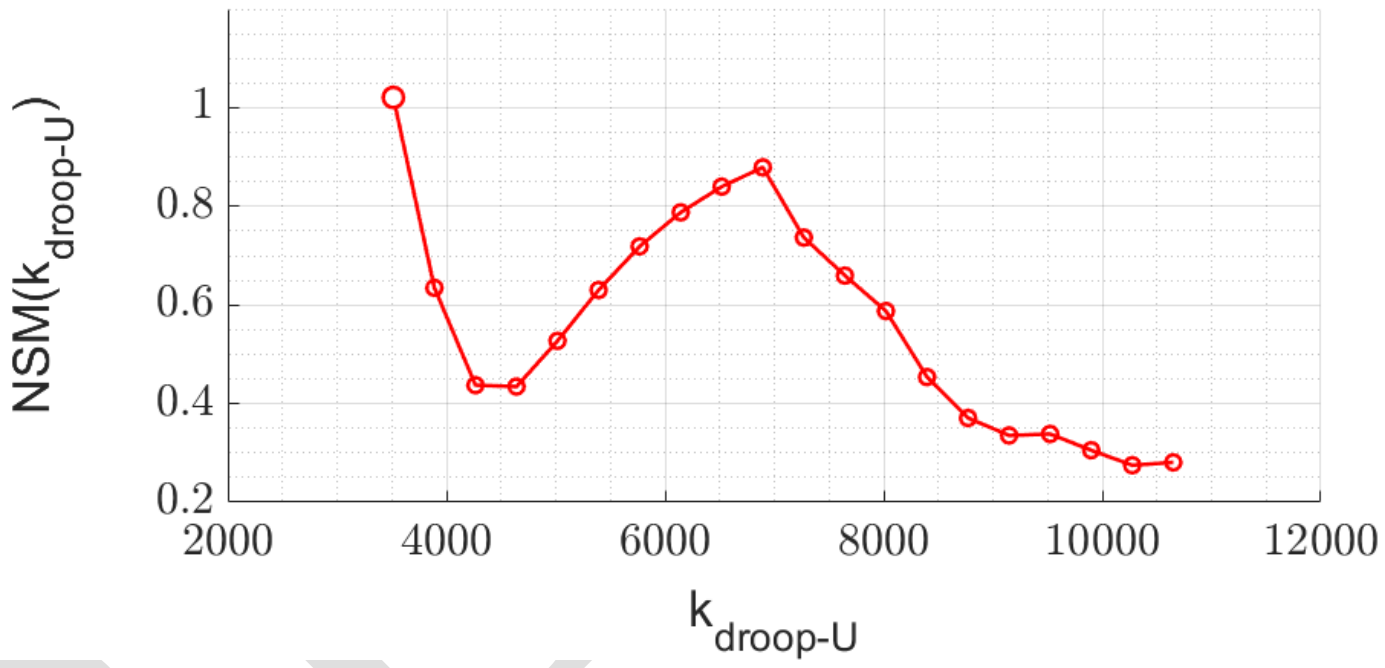

***Figure 12 – NSM index when varying the GFM $k_{droop\text{-}U}$ gain.***

Figure 13b illustrates the computed and normalized AEI, which is 17.91, smaller than the AEI obtained for the GFM. This confirms that the dominant mode experiences a smaller displacement per unit change in the droop gain, meaning that the GFL requires less adaptation effort. The effort of the GFL can also be tracked in the evolution of the derivatives of the dominant mode in Figure 13b, where, in contrast to Figure 10b, the variations are less observable and the only large change occurs at the 100% gain value. Finally, evaluating the local stability condition provided by (8) using the AEI shows that the only unstable operating point is the 100% gain case, matching the results obtained by the modal analysis in Figure 11. Note that, for the AEI, since the dominant modes presented for the GFM and GFL are quite similar, the normalization yields values around 18 for both cases; however, this may differ for other dominant modes, so the threshold is not fixed or standardized and should instead be compared across parameter variations to determine which system experiences less adaptation effort. When combined with the previous GFM results and the corresponding NSM values, the analysis clearly indicates that the GFL exhibits greater flexibility and requires less reconfiguration to achieve system stability. This finding challenges the common assumption that "the newly connected device must adapt to the grid," demonstrating instead that, in modern black-box-dominated power systems, the most flexible and least intrusive solution can be identified through the AEI.

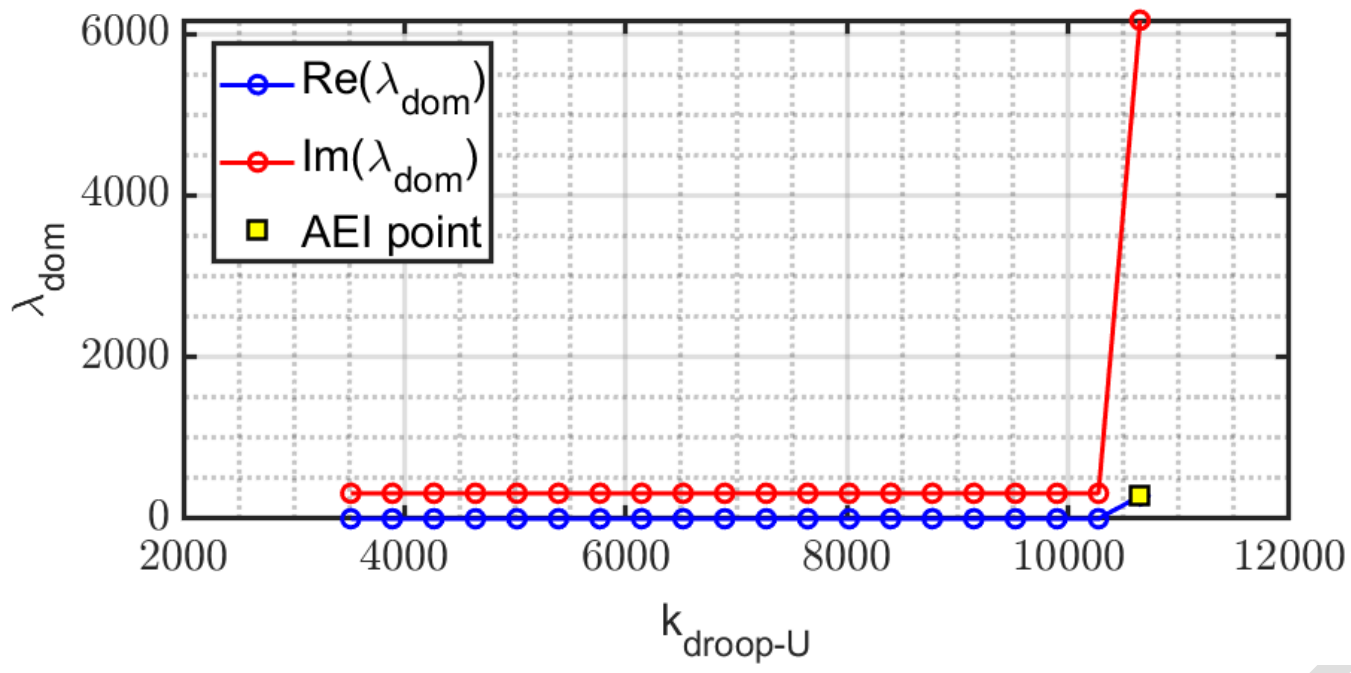

*a)*

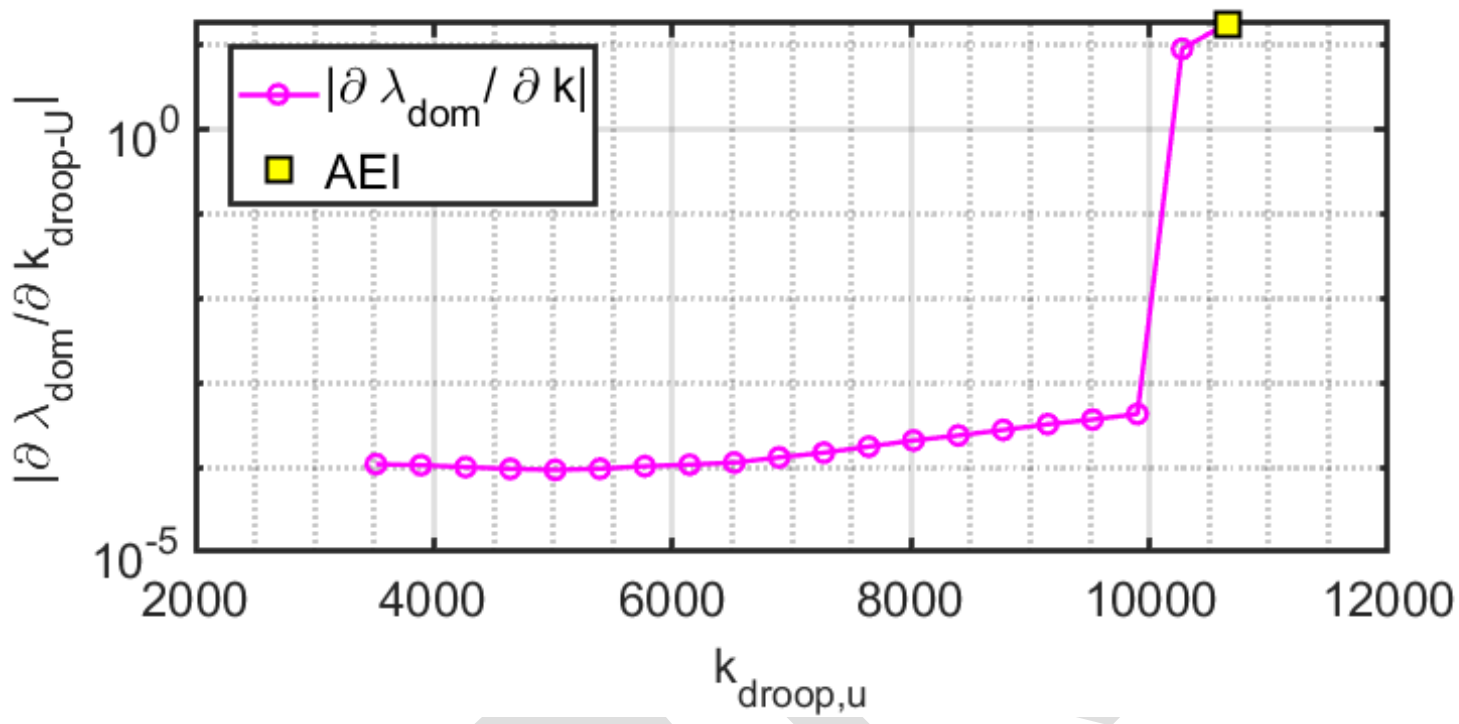

*b)*

***Figure 13 – Adaptation index results: a) dominant mode behaviour and b) AEI.***

# 4 Conclusions

This work introduced the Adaptation Effort Index (AEI), a metric designed to quantify how strongly a converter must adjust its control parameters to restore stability in black-box, converter-dominated power systems. By combining the Nyquist Stability Margin from frequency-domain screening with a modal-sensitivity index obtained through fitted state-space synthesis (fitss), the AEI provides a practical way to assess adaptability when detailed models are unavailable. Also, a local stability condition derived from the AEI is proposed as a compliance criterion for new grid-connection projects.

The method was applied to a system composed of a GFM and a GFL converter. Screening with the SIaD-Tool showed the stability limits originated from the reactive-power and voltage droop gains. Although reducing the GFM droop gain stabilizes the system, its AEI revealed that this adjustment requires a large modal displacement. In contrast, applying the same proportional reduction to the GFL gain restored stability with a much smaller AEI and a significantly higher NSM, indicating far lower adaptation effort.

Overall, the results show that the GFL converter provides the most flexible and least intrusive path to stability, challenging the assumption that newly connected devices must always adapt. The AEI thus offers a quantitative and scalable tool for identifying which converter should be reconfigured in modern black-box power systems and supplies a robust stability indicator for converter-based resource connection studies.

## 5 Acknowledgment

This work has received funding from the ADOreD project of the European Unions Horizon Europe Research and Innovation program under the Marie Skodowska-Curie grant agreement No. 101073554. The work of O. Gomis-Bellmunt and E. Prieto-Araujo was supported by the Agència de Gestió d'Ajuts Universitaris i de Recerca (AGAUR) through the ICREA Acadèmia programme, and by the Departament de Recerca i Universitats of the Generalitat de Catalunya. The work of E. Prieto-Araujo was also supported by the Serra Húnter Programme.

**APPENDIX**

**Algorithm 1.** Fitted State-Space (`fitss`) for Black-Box Models
**Input:** *Y(s)* or *Z(s)*, frequency points *f*, initial poles $p_0$, pole tolerance *tol*, reduction mode *red,* inputs name *in,* outputs name *out*
**Output:** Small-signal fitted state-space model *SSA-fitss*

1. Initialize poles $p_{\text{init}}$ ← initial poles $p_0$
2. Call `fitss`(*Y* or *Z, f,* $p_{\text{init}}$, *tol*, *red, in, out*)

**function** `fitss`(*Y* or *Z, f,* $p_{\text{init}}$, *tol*, *red*, *in*, *out*):

3. Validate input matrix dimensions of *Y* or *Z* and *f*
4. Set options *opts* and initialize vector fitting `vecfit3`
5. Extract *SISO* sub-vectors from *Y* or *Z*
6. Call `redfit`(*SISO*, $p_{\text{init}}$, *tol*, *red*)
7. Construct $MIMO_{fitted}$ state-space (A, B, C, D) using $SISO_{fitted}$
8. **while** system unstable or not converged **do**
    8.1 Update values $p_{\text{init}}$, *tol*, *red,* and restart fitting
    **end while**
9. Set *SSA-fitss* ← $MIMO_{fitted}$ *,in, out*
10. Return *SSA-fitss*

**function** `redfit`(*SISO*, $p_{\text{init}}$, *tol*, *red*):

11. **while** *RMS error* > pole tolerance *tol* and pole expansions allowed **do**
    11.1 Apply `vecfit3` to *SISO* sub-vectors
    11.2 Compute *RMS error* of fitted model
    11.3 **if** *RMS error* not improved **do**
        11.3.1 Insert new pole $p_{\text{new}}$ in max-error frequency region
        11.3.2 Update $p_{\text{init}}$ and pole expansions
        **end if**
    **end while**
12. Return $SISO_{full\text{-}fitted}$
13. Reduce order by singular value decomposition (SVD) or balanced realization (BR)
14. Output convergence diagnostics
15. Return $SISO_{fitted}$